\documentclass[aps,prb,twocolumn, showpacs]{revtex4}
\usepackage{amsmath,amssymb}
\usepackage{graphicx}
\usepackage{color}
\begin{document}

\title{Spin liquid like Raman signatures in hyperkagome iridate Na$_4$Ir$_3$O$_8$}
\author{Satyendra Nath Gupta$^1$, P. V. Sriluckshmy$^2$, Ashiwini Balodhi$^3$, D. V. S. Muthu$^1$, S. R. Hassan$^2$, Yogesh Singh$^3$, T. V. Ramakrishnan$^1$ and A. K. Sood$^1$}
\thanks{To whom correspondence should be addressed}
\email{asood@physics.iisc.ernet.in}
\affiliation{Department of Physics, Indian Institute of Science, Bangalore-560012,India
$^2$The Institute of Mathematical Sciences, C.I.T. Campus, Chennai 600 113, India
$^3$Indian Institute of Science Education and Research (IISER) Mohali, Knowledge City, Sector 81, Mohali 140306, India}      
 
\date{\today}

\begin{abstract}
Combining Raman scattering measurements with mean field calculations of the Raman response we show that Kitaev-like magnetic exchange is dominant in the hyperkagome iridate Na$_4$Ir$_3$O$_8$.  In the measurements we observe a broad Raman band at $\sim$~3500 cm$^{-1}$ with a band-width $\sim 1700$~cm$^{-1}$.  Calculations of the Raman response of the Kitaev-Heisenberg model on the hyperkagome lattice shows that the experimental observations are consistent with calculated Raman response where Kitaev exchange interaction ($J_K$) is much larger than the Heisenberg term J$_1$ ($J_1/J_K \sim 0.1$).  A comparison with the theoretical model gives an estimate of the Kitaev exchange interaction parameter. 
\end{abstract}
\pacs{78.30.Am, 75.10.Kt, 75.10.Jm, 78.20.Ls}

\maketitle
\section{Introduction}
Mott insulators with strong spin-orbit coupling can realize novel types of magnetic exchange and low energy Hamiltonians \cite{Jackeli2009, Chaloupka2010, Pesin2010, Wan2011}.  It was shown by Jackeli and Khaliullin\cite{Jackeli2009} that in materials with strongly spin-orbit entangled effective moments, the low energy effective magnetic Hamiltonians would depend on the lattice geometry and could interpolate between purely isotropic Heisenberg-like for corner shared octahedra with a $180^0$ transition metal-oxygen-transition metal (TM--O--TM) bond, to a bond-dependent quantum compass model for edge shared octahedra with a $90^0$ TM--O--TM bond.  For the specific case of a honeycomb lattice, the quantum compass model becomes the Kitaev model.  The Kitaev model is one of the simplest Hamiltonians for spins $S = 1/2$ on a honeycomb lattice which involves bond-dependent nearest neighbor interactions, is exactly solvable, and harbors a quantum spin liquid ground (QSL) state with Majorana Fermion excitations \cite{Kitaev2006}.  The suggestion that the Kitaev Hamiltonian and the related Kitaev-Heisenberg Hamiltonian \cite{Kitaev2006, Jackeli2009, Chaloupka2010} could be realized in a family of honeycomb lattice iridates $A_2$IrO$_3$ ($A =$~Na, Li)) has led to a flurry of activity on these materials \cite{Shitade2009,Singh2010,Singh2012,Choi2012,Ye2012,Gretarson2013, Comin2012, Kimchi2011,Chaloupka2013,Manni2014a,Katukuri2014,Rau2014} as well as recent work on the honeycomb lattice ruthenate $\alpha$--RuCl$_3$ \cite{Plumb2014, Majumder2015, Kubota2015, Shankar2015, Sears2015, Sandilands2016, Banerjee2016}.  While the spin-liquid state expected in the strong Kitaev limit has not been found experimentally for $A_2$IrO$_3$ or for $\alpha$--RuCl$_3$ there has been recent experimental work demonstrating the presence of dominant bond-dependent magnetic exchange and spin-space and real-space locking in Na$_2$IrO$_3$ \cite{Chun2015}, both of which are direct consequences of the presence of Kitaev-like magnetic exchange.  Additionally, Raman scattering measurements on Na$_2$IrO$_3$ \cite{Gupta2016}, Li$_2$IrO$_3$ \cite{Glamazda} and $\alpha$--RuCl$_3$ \cite{Sandilands2015} have revealed a broad, quasi-continuous polarization independent response similar to that predicted for the Kitaev spin liquid \cite{Knolle2014}.  The presence of such a feature in these magnetically ordered materials was interpreted as evidence for proximity to the Kitaev spin liquid \cite{Gupta2016, Glamazda, Sandilands2015}.  
 
The novel magnetic properties of these honeycomb lattice iridates and ruthenate most likely arise from the presence of dominant Kitaev-like interactions in competition with other residual Heisenberg-like or further neighbour interactions.  We recall that bond dependent interactions arise due to the strong spin-orbit coupling and edge shared TO$_6$ ($T =$~transition metal) octahedral geometry.  However, this geometry is common to several other structures and in particular is found in pyrochlore, spinel, and hyperkagome lattices.  Interestingly, iridate compounds are known for each of these structures: $R_2$Ir$_2$O$_7$, CuIr$_2$S$_4$, and Na$_4$Ir$_3$O$_8$.       

This work focuses on the hyperkagome iridate Na$_4$Ir$_3$O$_8$ which is a candidate for 3-dimensional quantum spin liquid \cite{Okamoto2007,Singh2013}.  No long ranged magnetic order was found\cite{Okamoto2007, Singh2013, Balodhi2015} down to $100$~mK despite strong antiferromagnetic exchange ($\theta \sim -600$~K) between effective spins $S = 1/2$.  Magnetic irreversibility below $6$~K hints at a glassy state \cite{Balodhi2015} which is confirmed by $\mu$SR and neutron diffraction measurements \cite{Dally2014} and by $^{23}$Na and $^{17}$O NMR measurements \cite{Shockley2015}.  However, properties above the freezing temperature are consistent with a spin liquid state \cite{Shockley2015}.  Whether the spin-glassy state is a result of disorder or due to several competing magnetic states is yet to be ascertained.  The ground state of ideal Na$_4$Ir$_3$O$_8$ samples may yet be a QSL.  

There have been several attempts\cite{Hopkinson2007, Lawler2008a, Zhou2008, Lawler2008b, Chen2008, Podolsky2009} to arrive at a minimal spin model that would best describe Na$_4$Ir$_3$O$_8$.  In most of these works predominantly the Heisenberg model on  the hyper-kagome lattice has been explored.  A recent study\cite{Kimchi2014} has explored the Kitaev-Heisenberg model on various lattices with edge shared octahedra including the hyperkagome lattice relevant for Na$_4$Ir$_3$O$_8$.  It is found that while the Kitaev spin-liquid exact solution doesn't generalize to this lattice, a quantum phase with extensive degeneracy is found in both limits of strong Kitaev or strong Heisenberg, with a 3D stripy order in between \cite{Kimchi2014}.  The stripy magnetic order has clearly not been found in experiments on Na$_4$Ir$_3$O$_8$. However, most thermodynamic measurements suggest proximity to a spin liquid state.  Which limit (Kitaev or Heisenberg) is more appropriate for the real material is thus still an open question.

We have measured the Raman response of high quality polycrystalline pellet samples of Na$_4$Ir$_3$O$_8$.  In addition to first order phonons we find a broad band with a maximum at $\sim$~3500 cm$^{-1}$ and with a band-width $\sim 1700$~cm$^{-1}$.  The broad band has some additional structure in contrast to the featureless response found earlier for Na$_2$IrO$_3$ \cite{Gupta2016}.  To understand these observations and to try to throw light on whether Heisenberg or Kitaev like interactions are dominant in Na$_4$Ir$_3$O$_8$ we have computed the Raman response for the nearest-neighbour Kitaev-Heisenberg model in both the strong Heisenberg and Kitaev limits.  The Raman response was calculated using the Majorana mean field framework assuming a spin liquid ground state for both limits. For the Heisenberg limit we find two peaks with relative intensity that does not match the experimentally observed Raman response. Even on introducing small Kitaev terms as perturbation does not give results which match our experiments.  For the pure Kitaev limit we obtain a broad band response.  There are however, additional features in the experiments which suggest the presence of other terms.  Hence we finally added small Heisenberg terms and find that additional peaks which develop, match the experimental observations.  Although the kitaev limit is not exactly solvable for the hyperkagome lattice we find a spin-liquid state for the parameters used to calculate the Raman response which match the experiments.  These results strongly indicate that Na$_4$Ir$_3$O$_8$ is a spin liquid close to dominant Kitaev limit with small Heisenberg perturbations.

\begin{figure}[t]
\includegraphics[width=0.4\textwidth]{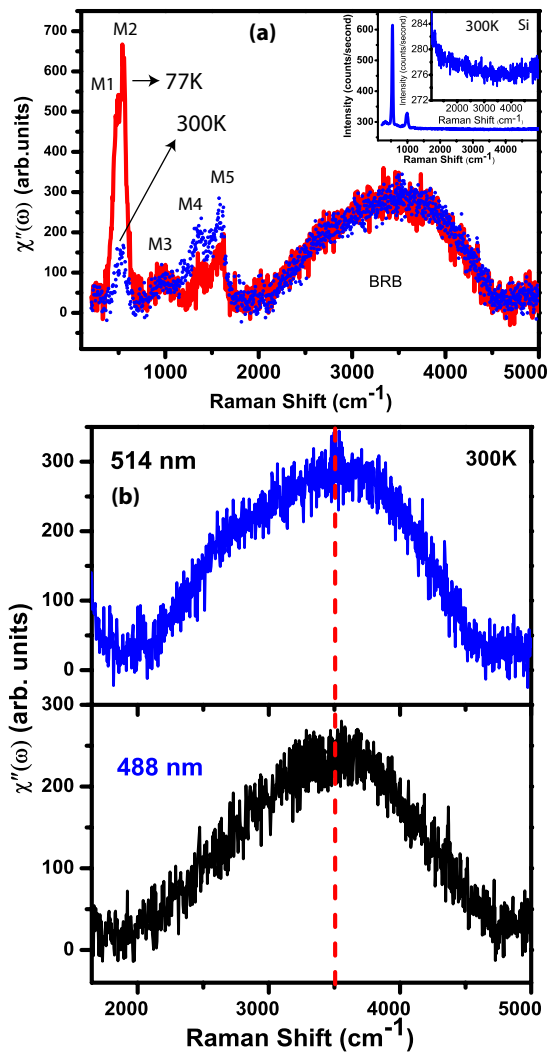}
\caption{(Color online)(a) Raman spectra of Na$_4$Ir$_3$O$_8$ measured at T = 77K (red line) and 300K (blue circles) in the spectral range 100 to 5000 cm$^{-1}$ using excitation laser wavelength of 514.5 nm. Inset: Raman spectra of silicon at 300K. The sharp lines near 520 cm$^{-1}$ and 1040 cm$^{-1}$ are first and second order Raman modes of Si respectively. The magnified Si spectra from 1000 to 5000 cm$^{-1}$ is shown in the inset. (b) Raman spectra recorded with two different laser excitation lines 514.5 and 488 nm. The vertical dashed line shows the center of the BRB. 
\label{Fig-Raman-77K}}
\end{figure}

\begin{figure}[b]
\includegraphics[width=0.5\textwidth]{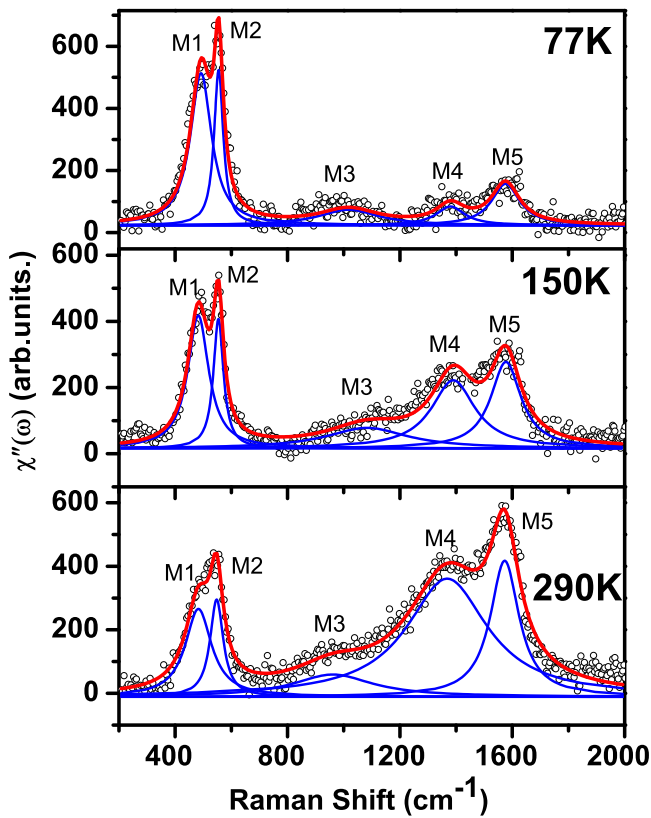}
\caption{Raman spectra of  Na$_4$Ir$_3$O$_8$ at three different temperatures 77K, 150K and 290K in the spectral range 200 to 2000  cm$^{-1}$. The solid blue lines are the Lorentzian fit to the individual peaks and solid red lines are sum of all the Lorentzians.   
\label{Fig-phonons}}
\end{figure}

\begin{figure}[t]
\includegraphics[width=0.5\textwidth]{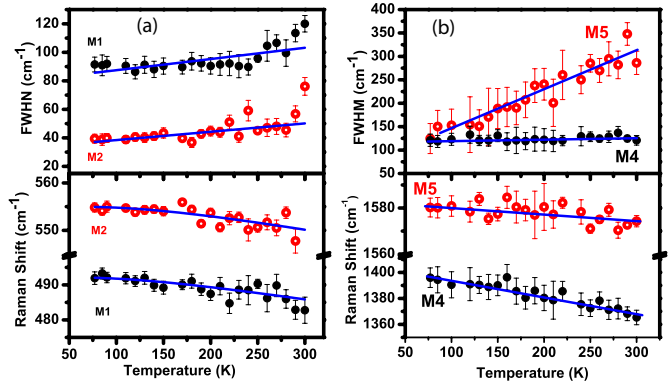}
\caption{(a)Temperature dependence of phonon frequency and FWHM of M1 (black filled circle) and M2 (red open circle) modes. The solid blue lines are fit to cubic anharmonic model. (b) Temperature dependence of phonon frequency and FWHM of M4 (black filled circle) and M5 (red open circle). The solid blue lines are guide to the eyes.
\label{Fig-BRB-Na2IrO3}}
\end{figure}
\section{Experimental Details}
Raman experiments are carried out on  polycrystalline pellets of Na$_4$Ir$_3$O$_8$. The synthesis and characterization of Na$_4$Ir$_3$O$_8$ have been reported elsewhere \cite{Singh2013, Balodhi2015}. The polycrystalline pellets  are polished to establish a virgin, optically flat surface for Raman measurements. Unpolarized micro-Raman measurements were performed  in backscattering geometry using 514.5 nm as well as 488 nm lines of an Ar-ion Laser  and  a confocal microscopy setup (WiTech) coupled with a Peltier cooled CCD. Temperature variation was done from 77K to 300K, with a temperature accuracy of ± 1K using continuous flow liquid nitrogen cryostat(Oxford Instrument). Spectra are recorded using a long working distance 50X objective with numerical aperture 0.45.

\section{Experimental Results} 
Na$_4$Ir$_3$O$_8$ has a cubic space group P4$_132$ with a unit cell containing four formula units ($60$ atoms), resulting in $180$ normal modes. According to factor group analysis, there are  $80~\Gamma$-point phonon modes out of which $44$ modes are Raman active.  Figure~\ref{Fig-Raman-77K} (a) shows the Raman susceptibility $\chi^{\prime\prime}$($\omega$) = Intensity($\omega$)/(n($\omega$)+1), where (n($\omega$)+1) is the Bose-Einstein factor, of Na$_4$Ir$_3$O$_8$ at $77$~K (red line) and 300K (blue circles) in the spectral range $100$ to $5000$~cm$^{-1}$, revealing $5$ Raman modes labeled as M1 to M5 and one broad Raman band centered at $3500$~cm$^{-1}$ abbreviated as BRB. It is evident from the figure that all the Raman modes M1 to M5 show temperature dependence while BRB is temperature independent. To rule out the possibility of instrumental artefacts as the origin of the BRB, we recorded Raman spectra of silicon at 300K up 5000 cm$^{-1}$ (inset of  Figure~\ref{Fig-Raman-77K} (a)).  The region from 1000 cm$^{-1}$ to 5000 cm$^{-1}$ has been magnified and shown in the inset in order to see any broad feature. It is clear from the inset  that there is no  feature present in the Raman spectra of Silicon and hence confirms that the BRB seen at  $\sim 3500$~cm$^{-1}$ is intrinsic to Na$_4$Ir$_3$O$_8$.   In order to rule out the possibility of photoluminescence as a cause for the origin of the broad band, Raman spectra recorded with a different laser line (488 nm) at 300K shows the same mode without any frequency shift as shown in Figure~\ref{Fig-Raman-77K} (b) and hence rules out the broad band to be related to photoluminescence. 
 
The modes M1 ($\sim{490 cm^{-1}}$) and M2 ($\sim{550 cm^{-1}}$) are first order Raman modes associated with the phonons. The mode M3 ($\sim{1000 cm^{-1}}$) could be a second order Raman mode (ie 2$\omega_{M1}$ or $\omega_{M1}+\omega_{M2}$) or a magnetic excitation. The exact assignment of M1, M2 and M3 will require full phonon calculations which is not yet reported.  At $300$~K the modes M4 ($\sim{1395 cm^{-1}}$) and M5 ($\sim{1580 cm^{-1}}$) are stronger than the mode M2 and hence cannot be higher order Raman phonon modes.  The temperature dependence of these two modes is also opposite to that expected for phonon modes.  We tentatively assign these to the magnetic excitations and return to it.

In order to estimate peak frequencies and full width at half maximum (FWHM) in the investigated temperature range, Lorentzian line shapes were used to fit the Raman modes M1 to M5.  Figure~\ref{Fig-phonons} shows the fitted  spectra collected at three different temperatures $77$~K, $150$~K and $290$~K.  Temperature evolution of phonon frequencies and FWHM of M1 and M2 modes are shown in Fig.~\ref{Fig-BRB-Na2IrO3} (a). The solid blue lines are the fit to  a simple cubic anharmonicity model where the phonon decays into two phonons of equal frequency\cite{Klemens1966}. It is clear from Fig.~\ref{Fig-BRB-Na2IrO3} (a) that  the modes M1 and M2 follow normal anharmonic behavior.    The phonon frequency and line-width of M3 mode do not show significant change with temperature and hence  not shown.  Figure~\ref{Fig-BRB-Na2IrO3}(b)  shows the temperature dependence of the peak frequencies and FWHM of M4 and M5 modes. The solid blue lines are the guide to the eye. The line-width of M4 mode is almost constant while the M5 mode broadens by $\sim 150 cm^{-1}$ with increasing temperature.

We now focus on the broand band response. Recent work on Li$_2$IrO$_3$ \cite{Glamazda} and $\alpha$--RuCl$_3$ \cite{Sandilands2015} (with $\left|\theta_{cw}\right|\sim40K$) show temperature dependence of the BRB. In comparison, the observed BRB for Na$_4$Ir$_3$O$_8$ at 3500 cm$^{-1}$ ($\approx 0.4$~eV) in the temperature range $T_N < T < \left|\theta_{cw}\right|$ does not show any temperature dependence because the Curie-Weiss temperature ($\left|\theta_{cw}\right|\sim650K$) is much larger compared to the temperature range 77K to 300K covered in our experiment.    The possibility of the BRB being a two magnon peak is unlikely since the system does not order magnetically.  Another possible origin of BRB can be electronic excitation from Ir 5d-shell multiplet, as seen recently in Sr$_2$IrO$_4$ \cite{Yang} where the temperature dependent Raman bands at $\sim{5600~ cm^{-1}}$ and $\sim{5450 ~cm^{-1}}$ are observed, in agreement with similar resonances seen in resonant inelastic x-ray scattering (RIXS)~ \cite{Kim 2012, Kim 2014}. In Na$_4$Ir$_3$O$_8$, RIXS spectrum \cite{Takayama}  shows  bands at $\sim 1 eV $ and 4 eV associated with the inter-atomic excitations within the d-orbitals of Ir. The temperature independence of the observed BRB and the absence of a similar energy scale in RIXS~ \cite{Takayama} rule out the assignment of the BRB to the electronic excitation within the 5d multiplets.

\begin{figure*}[t]
\includegraphics[width=0.7\textwidth]{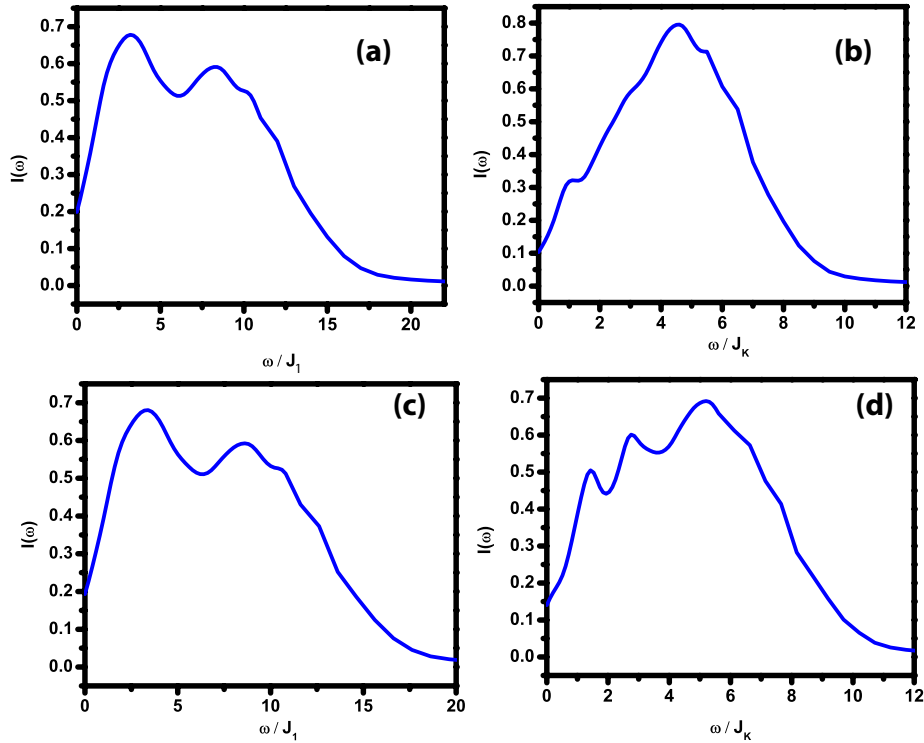}
\caption{(Color online) Theoretical curves: a) Pure Heisenberg model, b) Pure Kitaev model, c) Heisenberg model with small Kitaev interaction $J_K = 0.1$, $J_1 = 1$ with $J_K/J_1 = 0.1$ and d) Kitaev model with small Heisenberg interaction $J_K = 1.96$, $J_1 = 0.2$ with $J_1/J_K \sim 0.1$. The broadening used  in (a) and (c) is ~$\epsilon = 0.8J_1$ while in (b) and (d), it is $\epsilon = 0.4J_K$.
\label{Fig-Raman-300K}}
\end{figure*}

\begin{figure}[t]
\includegraphics[width=0.45\textwidth]{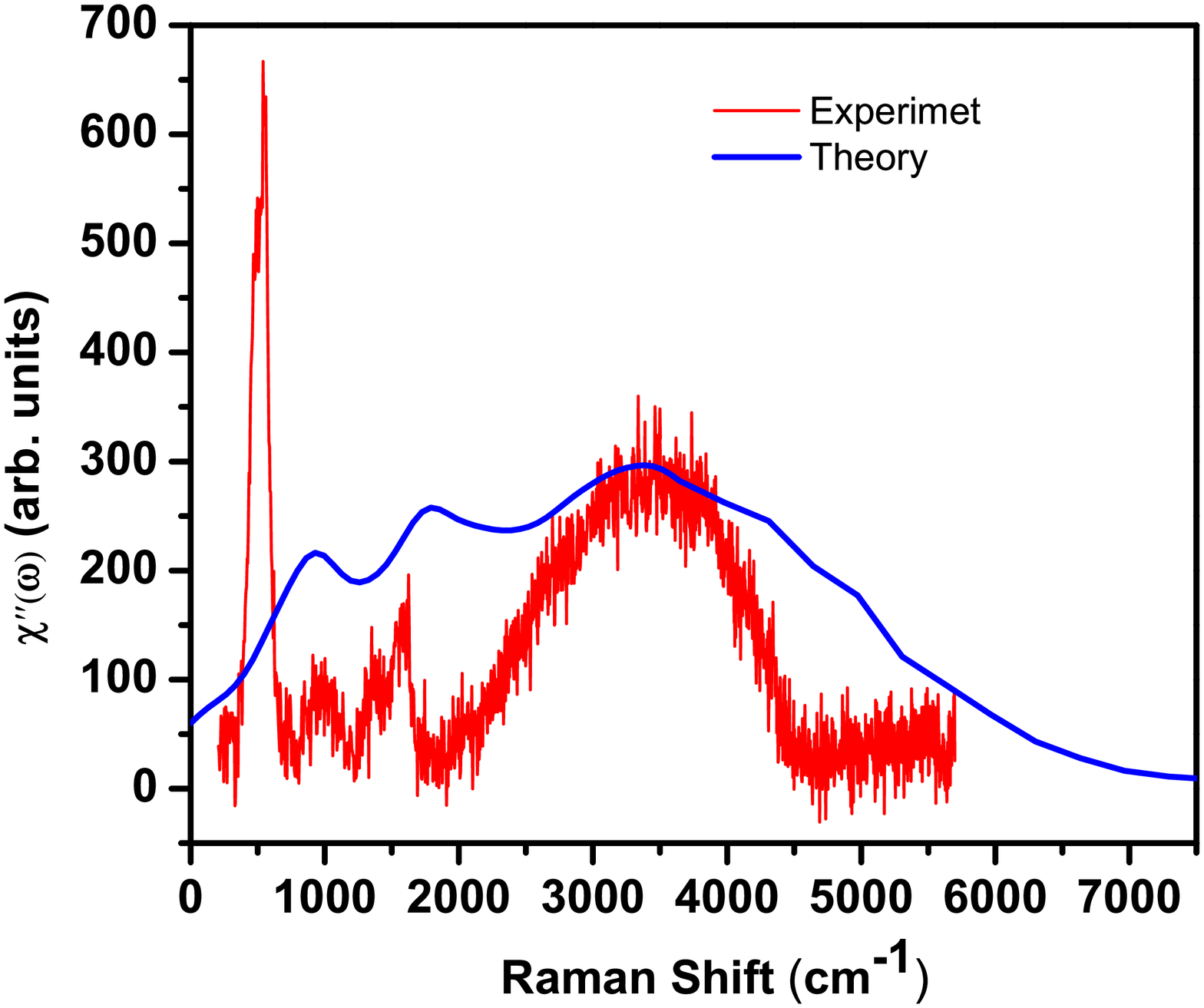}
\caption{(Color online) Comparison of experimental (red) and theoretical (blue) Raman spectrum. Here the frequency scale (J$_K$) in the calculation has been chosen (J$_K$=75meV) to match the main band position with the experiment. 
\label{Fig-experimenttheory}}
\end{figure}

Finally we consider a more interesting possibility.  The BRB could have a magnetic origin and be a signature of fractionalized excitations arising from a spin-liquid ground state.  Such BRB's seem to be a generic feature of spin liquids and have been predicted for the spin liquid state in Herbertsmithite \cite{Cepas2008, Ko2010} and for the Kitaev spin liquid on the honeycomb lattice \cite{Knolle2014}, for example.  The predicted BRB's have also been observed for the spin liquid candidate Herbertsmithite \cite{Wulferding2010},   Na$_2$IrO$_3$ \cite{Gupta2016},  Li$_2$IrO$_3$ \cite{Glamazda} and $\alpha$-RuCl$_3$ \cite{Sandilands2015}, in which Na$_2$IrO$_3$, Li$_2$IrO$_3$  and $\alpha$-RuCl$_3$ are candidate Kitaev materials.  To pursue this line we have calculated the Raman response of the Kitaev-Heisenberg model on the hyperkagome lattice within a  Majorana fermion based mean-field theory.  We have studied both the extreme limits of purely Heisenberg exchange (no Kitaev) and purely Kitaev exchange.  We have also studied the effect on the Raman response in these two limits of adding small perturbations of the other kind.

\section{Theoretical Calculations}
The Raman response obtained from our theoretical calculations is shown in Fig.~\ref{Fig-Raman-300K} (For details see Supplementary material). For the exact Kitaev spin liquid ground state on the honeycomb lattice, the Raman spectrum  was shown to be a broad polarization independent band, essentially due the propagating Majorana fermions \cite{Knolle2014}.  We have studied both Heisenberg and Kitaev limits for the hyperkagome lattice assuming a spin liquid ground state.  The calculated Raman response for these two cases are shown in Figs.~\ref{Fig-Raman-300K}~(a) (pure Heisenberg) and (b) (pure Kitaev).  The Raman response of the pure (antiferomagnetic) Heisenberg limit shows a two peak structure arising due to the spinon and gauge sectors, with the lower energy peak being more intense, very different from the experimentally observed BRB in Fig.~\ref{Fig-Raman-77K}. On introducing small Kitaev perturbations the curves  do not vary much as shown in Fig.~\ref{Fig-Raman-300K}~(c). The calculated Raman response of the pure Kitaev model (Fig.~\ref{Fig-Raman-300K}~(b)) reveals a broad band similar to the experiments, but there are additional peaks (M3, M4 and M5 modes) in the experimental data which need to be explained. On the addition of a small Heisenberg term ($J_1/J_K = 0.1$) as a perturbation to the Kitaev term we obtain a response shown in Fig.~\ref{Fig-Raman-300K}~(d) which looks a better match to the experimentally observed BRB. The calculated BRB is broad and has additional weak features at lower energies. The theoretical Raman response for $J_1/J_K = 0.1$ shown in Fig.~\ref{Fig-Raman-300K}~(d) is plotted with the experimental curve for comparison in Fig.~\ref{Fig-experimenttheory}. We note that the calculated Raman response is broader than the observed lineshape, perhaps due to mean field calculations used. It is clear that theoretical Raman response calculated with small Heisenberg term ($J_1/J_K = 0.1$) as a perturbation to the Kitaev term has a better match with experimental data.    Thus   the Raman response calculated for the pure Heisenberg limit is inconsistent with our observed BRB while the strong Kitaev limit with small Heisenberg term gives results consistent with experiments. The comparison of experimental and theoretical data (Fig.~\ref{Fig-experimenttheory}) gives the estimate of Kitaev intraction to be  $J_K \sim 75$~meV. This value is  high but  is consistent with the  large Weiss temperature of $-650$~K obtained from magnetic measurements \cite{Okamoto2007, Balodhi2015}. Note that the two additional weak features at  920 cm$^{-1}$ and 1650 cm$^{-1}$ in the calculated BRB   are close to the experimentally observed M3 (1000 cm$^{-1}$), M4 and M5 ($\sim$1580 cm$^{-1}$) modes  (see Fig.~\ref{Fig-experimenttheory}). The mode M3 may not be a second order phonon mode.

\section{Conclusions}
In conclusion, we have experimentally shown the existence of a broad Raman band at high energies for Na$_4$Ir$_3$O$_8$.  By calculating the Raman response for the Kitaev-Heisenberg model on the hyperkagome lattice we show that the observed BRB is in very good agreement with calculated Raman response for the Kitaev limit with small Heisenberg perturbations ($J_1/J_K= 0.1$).  Although the Kitaev limit is not exactly solvable for the hyperkagome lattice we find a spin-liquid state for the parameters used to calculate the Raman response which match the experiments. This strongly suggests that Na$_4$Ir$_3$O$_8$ is a spin-liquid driven by strong Kitaev interactions with smaller Heisenberg terms.  

\section*{ACKNOWLEDGMENTS}
A.K.S. and T. V. R. acknowledge funding from DST.  YS acknowledges partial support from DST through the Ramanujan fellowship and through the grant no. SB/S2/CMP-001/2013.

\clearpage

\renewcommand\thefigure{S\arabic{figure}}
\begin{center}
\textbf{Supplementary Material}
\end{center}

The Kitaev-Heisenberg model on the hyper-kagome lattice can be written as
\begin{align}
\mathcal{H} & = \sum_{\langle ij\rangle^\alpha} \left[ J_{k} S_i^\alpha S_j^\alpha + J_1 {\bf S}_i \cdot {\bf S}_j\right] . \label{khhyperk}
\end{align}
\begin{figure}[b]
\includegraphics[width=0.45\textwidth]{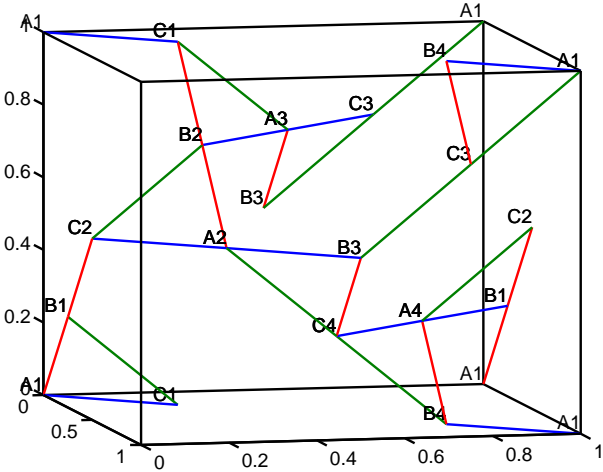}
\caption{Kitaev model on the hyper-kagome lattice showing the $12$ site unit cell with the colors corresponding to the $x,y,z$ interactions.
\label{fig:kitaev_hyperk}}
\end{figure}
Fig.\ref{fig:kitaev_hyperk} shows a pure Kitaev model on the hyperkagome unit cell with $\alpha = x,y,z$ links, color coded as red, blue and green. Theoretically, the challenge of solving the Hamiltonian is enhanced by the large number of basis sites(12) in the hyper-kagome lattice. This makes it difficult to effectively apply any numerical technique. Lawler et al. \cite{ Lawler2008b} worked on the Heisenberg model using fermionic mean field theory and showed that the $Z_2$ spin liquid is energiticaly favoured. On the other hand the Kitaev model which gives an exactly solvable $Z_2$ spin liquid ground state on the honeycomb lattice, is not exactly solvable on the hyper-kagome lattice \cite{Kimchi2014}. Thus expecting an isotropic spin liquid ground state in both the limits, the above Hamiltonian is written in terms of Majorana fermions\cite{Gupta2014}. 
\begin{align}
\sigma_i^\alpha\sigma_j^\beta=-ic_ic_j ~ ib^\alpha_ib^\beta_j
&\approx -ic_i c_j B^{\alpha\beta}_{ij}-iC_{ij}b_i^\alpha b_j^\beta+ C_{ij} B^{\alpha\beta}_{ij}.
\end{align}
where the mean field of these operators, $c$ spinon and $b^\alpha$ vison sector, involve the nearest neighbour Majorana correlations alone. The Spinon and Vison Hamiltonians decouple, effectively affecting the hopping of the other. The self-consistency mean field parametric equations are

\setlength{\abovedisplayskip}{0pt}%
\begin{align}
B^{\alpha\beta}_{\langle ij\rangle^\gamma}\equiv\langle ib^\alpha_ib^\beta_{j^\gamma}\rangle \equiv B^{\alpha\beta}_{\gamma} \qquad\qquad
C_{\langle ij\rangle^\gamma}\equiv\langle ic_ic_{j^\gamma}\rangle \equiv C_{\gamma} .\label{params}
\end{align}

\begin{figure}[t]
\includegraphics[width=0.45\textwidth]{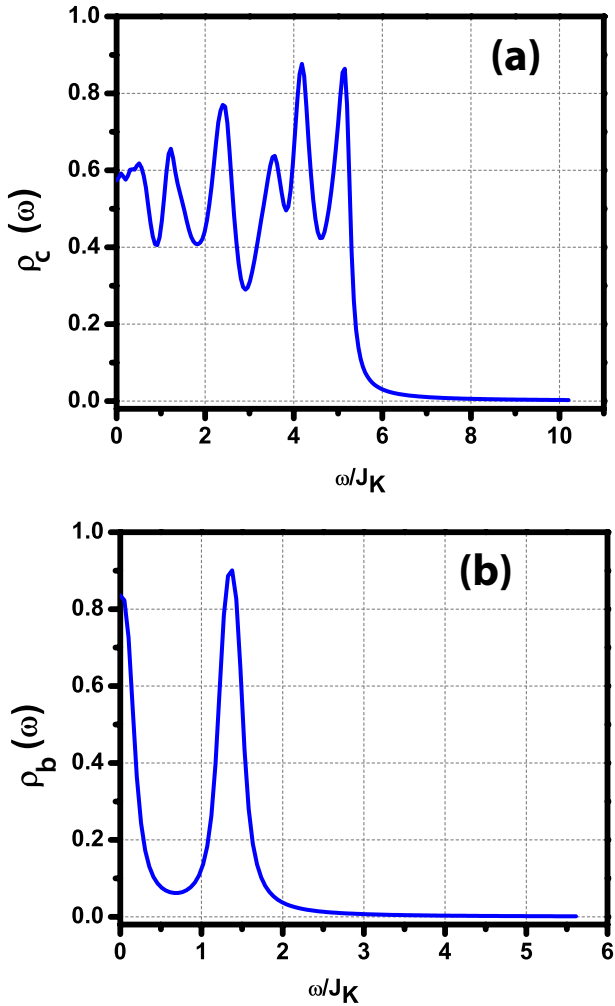}
\caption{(Color online) Density of states for the Spinon (a) and Visons (b) for the Kitaev model with small Heisenberg interaction $J_k = 1.96$, $J_1 = 0.2$ with $J_1/J_k \sim 0.1$. 
\label{fig:dos}}
\end{figure}

The mean field Hamiltonian written in momentum basis $\mu_k$ and $\nu_k^\alpha$ becomes
\begin{align}
H_{MF}&  = H_c + H_b + \sum_{\langle ij\rangle^\alpha} \left[ J_{k}  C_{\alpha}B^{\alpha\alpha}_{\alpha}  + J_1 \sum_{\beta} C_{\alpha}B^{\beta\beta}_{\alpha}  \right] \\
H_c&  = \sum_{\langle ij\rangle^\alpha} \left[ J_{k} \left[-ic_ic_jB^{\alpha\alpha}_{\alpha}\right] + J_1 \sum_{\beta} \left[-ic_ic_jB^{\beta\beta}_{\alpha} \right] \right]\nonumber\\
&=\sum_k \mu_k^\dagger ~ h_k^c ~\mu_k\\
H_b & = \sum_{\langle ij\rangle^\alpha} \left[ J_{k} \left[-iC_{\alpha}b_i^\alpha b_j^\alpha \right]+ J_1 \sum_{\beta} \left[-iC_{\alpha}b_i^\beta b_j^\beta \right] \right]\nonumber\\
& = \sum_{k\alpha} \nu_k^{\alpha\dagger}~ h_k^{b^\alpha}~ \nu_k^\alpha
\end{align}
Let $M_k$ and $N_k^{\alpha}$ represent the unitary matrices that diagonalize $h_k^c$ and $h_k^{b^\alpha}$ respectively giving the simplified Hamiltonians

\begin{align}
\mathcal{H}_c & = \sum_k \mu_k^\dagger M_k ~(M_k^\dagger h_k^c M_k)~ M_k^\dagger \mu_k\nonumber\\
 &= \sum_k \left(f_k^\dagger, ~ f_{-k}\right)\left(\begin{array}{cc}
E_k & 0 \\
0 & -E_k
\end{array} \right) \left(\begin{array}{c}
f_k\\
f_{-k}^\dagger
\end{array}\right)\\
&  = \sum_k E_k^a (2f_k^{a\dagger} f_k^a - 1)  \\
\mathcal{H}_b & = \sum_{k\alpha} \nu_k^{\alpha\dagger} N_k^\alpha ~ ( N_k^{\alpha\dagger} h_k^{b^\alpha} N_k^\alpha )~ N_k^{\alpha\dagger} \nu_k^\alpha \nonumber\\
&= \sum_k \left(g_k^{\alpha\dagger}, ~ g_{-k}^\alpha\right)\left(\begin{array}{cc}
E_k^\alpha & 0 \\
0 & -E_k^\alpha
\end{array} \right) \left(\begin{array}{c}
g_k^\alpha\\
g_{-k}^{\alpha\dagger}
\end{array}\right) \\
& = \sum_{k\alpha} E_k^{a\alpha} (2g_k^{a\alpha\dagger} g_k^{a\alpha} - 1)  
\end{align}
where $a = 1,2, \cdots 6$ and $\alpha = x,y,z$. The diagonal operators $f_k$ and $g_k^\alpha$ and the eigenvalues $E_k$, $E_k^\alpha$ can be computed numerically. The single particle density of states for the spinons as shown in Fig.\ref{fig:dos}(a) has a lot of features. On the other hand, the vison density of states as shown in Fig.\ref{fig:dos}(b) has two peaks one centered around $\omega_{p1} = 0$ and the other around $\omega_{p2} = 1.5 J_k$. The ground state is $|GS\rangle = \Pi_{k,a,b,\alpha} f^{a}({\bf k})g^{b\alpha} ({\bf k}) |0\rangle$.  These peaks play an important role in the Raman response of the system. 
The operators evolve as

\begin{align}
f_k^a(t) & = f_k^a (0) e^{-2iE_k^a t}; \qquad f^{a\dagger}_k(t) = f^{a\dagger}_k (0) e^{2iE_k^at} \label{timeevolf}\\
g_k^{a\alpha}(t) & = g_k^{a\alpha} (0) e^{-2iE_k^{a\alpha} t}; \qquad g^{a\alpha\dagger}_k(t) = g^{a\alpha\dagger}_k (0) e^{2iE_k^{a\alpha}t} \label{timeevolg}
\end{align}

In this paper, our focus is to understand the material based on Raman Intensity experiments. For the mean field ground state, the Raman Intensity is computed \cite{Gupta2014} 
\begin{align}
I(\omega) & = \int ~dt~ e^{i\omega t}~ iF(t) = \int ~dt~ e^{i\omega t}~\langle GS|R(t)R(0)|GS\rangle \label{rr}
\end{align}
where the Raman operator is given by\cite{Knolle2014}
\begin{align}
R & =  \sum_{\langle ij \rangle \alpha} ({\boldsymbol{\epsilon}}_{in}\cdot{\boldsymbol{d}}^\alpha) ({\boldsymbol{\epsilon}}_{out}\cdot{\boldsymbol{d}}^\alpha) (K S_i^\alpha S_j^\alpha + K_1  {\bf{S}}_i \cdot {\bf{S}}_j) \nonumber\\
&= \sum_{\langle ij \rangle \alpha} m_\alpha (K S_i^\alpha S_j^\alpha + K_1  {\bf{S}}_i \cdot {\bf{S}}_j)
\end{align}
$K\propto J_K$, $K_1\propto J_H$, ${\boldsymbol{\epsilon}}_{in/out}$ correspond to the incoming and outgoing polarization directions respectively and ${\boldsymbol{d}}^\alpha$ the nearest neighbour bond vectors. The calculation for the response is illustrated for the pure Kitaev model as


\begin{align}
iF(t) & = K^2 \sum_{\langle ij \rangle \alpha} \sum_{\langle kl \rangle \beta} m_\alpha m_\beta \langle ic_i(t) c_j(t) ib_i^\alpha (t) b_j^\alpha (t) ic_k(0) c_l(0)\nonumber \\&~~~~~~~~~~~~~~~~~~~~~~~~~~~~~~~~~~~~~~~~~~~~~~~~~~~~~~
 ib_k^\beta (0) b_l^\beta (0) \rangle \\
& = K^2 \sum_{\langle ij \rangle \alpha} \sum_{\langle kl \rangle \beta} m_\alpha m_\beta \langle ic_i(t) c_j(t) ic_k(0) c_l(0)\rangle\nonumber \\&~~~~~~~~~~~~~~~~~~~~~~~~~~~~~~~~~
\langle ib_i^\alpha (t) b_j^\alpha (t) ib_k^\beta (0) b_l^\beta (0) \rangle
\end{align}

\begin{figure*}[b]
\includegraphics[width=0.8\textwidth]{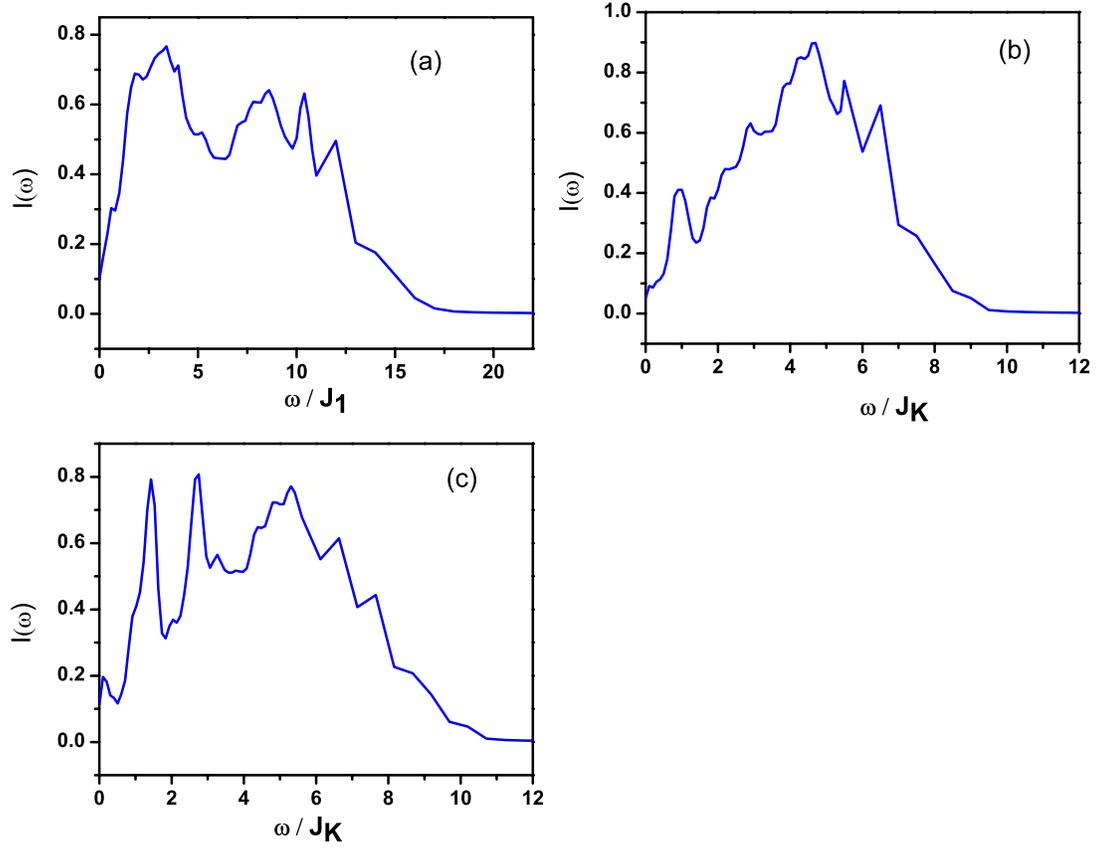}
\caption{(Color online) Theoretical curves: a) Pure Heisenberg model, b) Pure Kitaev model and c) Kitaev model with small Heisenberg interaction $J_k = 1.96$, $J_1 = 0.2$ with $J_1/J_k \sim 0.1$. The broadening used  in (a) is ~$\epsilon = 0.2J_1$ while in (b) and (c), it is $\epsilon = 0.1J_K$.
\label{Fig-unbroadened}}
\end{figure*}

In the spinon sector a correlation function can be expanded 
\begin{align}
\langle ic_i(t) c_j(t)  ic_k(0) c_l(0)\rangle & = \langle ic_{i}(t) c_{j}(t) \rangle\langle ic_{k}(0) c_{l}(0)\rangle  \nonumber\\
&- \langle ic_{i}(t)   c_{k}(0)\rangle\langle ic_{j}(t) c_{l}(0)\rangle \nonumber \\ 
& + \langle ic_{i}(t)c_{l}(0) \rangle\langle i c_{j}(t)  c_{k}(0) \rangle 
\end{align}
From Eq:\eqref{params} the first term becomes a constant
\begin{align}
\langle ic_{i}(t) c_{j}(t) \rangle = C_{\gamma}; \qquad \langle ic_{k}(0) c_{l}(0) \rangle = C_{\gamma'}.
\end{align}
Similar expressions for the vison sector can be obtained. Considering only the dominant time dependant contribution to the Raman intensity, the Raman operator becomes
\begin{align}
iF(t) & \approx K^2 \sum_{\langle ij \rangle \alpha} \sum_{\langle kl \rangle \beta} m_\alpha m_\beta \langle ic_i(t) c_j(t)  ic_k(0) c_l(0)\rangle  \nonumber B^{\alpha\alpha}_{\alpha} B^{\beta\beta}_{\beta} \\
& + K^2 \sum_{\langle ij \rangle \alpha} \sum_{\langle kl \rangle \beta} m_\alpha m_\beta C_\alpha C_\beta \langle ib_i^\alpha (t) b_j^\alpha (t) ib_k^\beta (0) b_l^\beta (0) \rangle \\
& = \langle R_c(t)R_c(0)\rangle  + \langle R_b(t)R_b(0)\rangle 
\end{align}
where $R_c$ and $R_b$ are the Raman operators in the spinons and visons alone and rewritten in the diagonal operators $f({\bf k}),g^{\alpha}({\bf k})$ as
\begin{align}
R_c & = K \sum_{\langle ij \rangle \alpha} m_\alpha  ic_i c_j  B^{\alpha\alpha}_{\alpha} = \sum_k \mu_k^\dagger \tilde{h_k} \mu_k \nonumber\\
 &= \sum_k \mu_k^\dagger M_k (M_k^\dagger \tilde{h_k} M_k) M_k^\dagger \mu_k \\
& = \sum_k \left(f_k^\dagger, ~ f_{-k}\right)\left(\begin{array}{cc}
F_{1}({\bf{k}}) & F_{2}({\bf{k}}) \\
F_2^\dagger ({\bf{k}}) & -F_1({\bf{k}})
\end{array} \right) \left(\begin{array}{c}
f_k\\
f_{-k}^\dagger
\end{array}\right)\\
& =\sum_k F_1^{ab}({\bf{k}}) f_k^{a\dagger} f_k^{b} + F_2^{ab} ({\bf k}) f_k^{a\dagger} f^{b\dagger}_{-k} + h.c. \\
R_b & =  K \sum_{\langle ij \rangle \alpha} m_\alpha  C_\alpha ib_i^\alpha  b_j^\alpha = \sum_{k\alpha} \nu_k^{\alpha\dagger} \tilde{h_k^\alpha} \nu_k^\alpha\nonumber\\
&= \sum_{k\alpha} \nu_k^{\alpha\dagger}N_k^\alpha ( N_k^{\alpha\dagger}  \tilde{h_k^\alpha}N_k^\alpha ) N_k^{\alpha\dagger}  \nu_k^\alpha \\
& = \sum_{k\alpha} \left(g_k^{\alpha\dagger}, ~ g_{-k}^\alpha\right)\left(\begin{array}{cc}
G^\alpha_{1}({\bf{k}}) & G_{2}^\alpha({\bf{k}}) \\
G_2^{\alpha\dagger} ({\bf{k}}) & -G_1^\alpha({\bf{k}})
\end{array} \right) \left(\begin{array}{c}
g_k^\alpha\\
g_{-k}^{\alpha\dagger}
\end{array}\right)\\
& = \sum_{k\alpha} G_1^{ab\alpha}({\bf{k}}) g_k^{a\alpha\dagger} g_k^{b\alpha} + G_2^{ab\alpha} ({\bf k}) g_k^{a\alpha\dagger} g^{b\alpha\dagger}_{-k}   + h.c.
\end{align}


with the time evolution Eq:\eqref{timeevolf},\eqref{timeevolg}. Therefore the dominant contribution to the intensity can be written as
\begin{align}
I(\omega) & \approx   2\pi \sum_{{\bf{k}},a,b} \delta(\omega - 2E^a({\bf{k}}) - 2E^b({\bf{k}})) ~|F_2^{ab}({\bf k})|^2  \nonumber \\
& + 2\pi \sum_{{\bf{k}},a,b,\alpha}\delta(\omega - 2E^{a\alpha}({\bf{k}}) - 2E^{b\alpha}({\bf{k}})) ~|G_2^{ab\alpha}({\bf k})|^2  \\
& = - 2 {\it{Im}} \left[ \sum_{{\bf{k}},a,b} \frac{1}{\omega - 2E^a({\bf{k}}) - 2E^b({\bf{k}}) + i\epsilon}~ |F_2^{ab}({\bf k})|^2 \right] \nonumber \\
& - 2 {\it{Im}} \left[ \sum_{{\bf{k}},a,b,\alpha} \frac{1}{\omega - 2E^{a\alpha}({\bf{k}}) - 2E^{b\alpha}({\bf{k}}) + i\epsilon}~ |G_2^{ab\alpha}({\bf k})|^2\right]
\end{align}

where $\epsilon$ is the broadening parameter. The calculation can be easily extended to the Kitaev-Heisenberg model. The Raman response computed for the broadening parameter ~$\epsilon = 0.8J_1$ for pure Heisenberg , $\epsilon = 0.4J_K$ for pure Kitaev and both Kitaev and Heisenberg are shown in the paper. For less broadening ~$\epsilon = 0.2J_1$ for pure Heisenberg , $\epsilon = 0.1J_K$ for pure Kitaev and both Kitaev and Heisenberg, the computed Raman response is given in Fig.\ref{Fig-unbroadened}. The wiggles in the Raman response at lower broadening stems from those present in the spinon density of states shown in Fig.\ref{fig:dos}(a). The two sharp peaks in Fig.\ref{Fig-unbroadened}(c) occuring around $\omega = 1.5 J_k$ and $\omega = 2.7J_k$, arises from the peaks in the vison density of states: around $\omega_{p2}$ and $2\omega_{p2}$. The Raman Response from the theoretical calculations is found to be polarization independant which is a feature of the spin-liquid ground state.

\end{document}